# The current state of open access

Shigeki SUGITA (Kyoto University Library)

From the standpoint of a university librarian who has been involved in the establishment and administration of institutional repositories, I would like to offer a personal reflection. The term *"open access"* itself may be somewhat misleading. The phrase tends to evoke the image of making access open—that is, enabling anyone to read the material. This frames the issue primarily from the perspective of the reader. Yet this is not the true origin of the open access movement. Its foundations lay in the author's own desire to disseminate scientific knowledge to the world. Perhaps it would have been better had we begun with terms such as *open publishing* or *open sharing*.

There is a well-known painting by Raffaello Sanzio—*The School of Athens* (*Scuola di Atene*)—depicting eminent philosophers and scientists of ancient Greece: Socrates, Plato, Aristotle. The scene brings together figures who lived in different eras, and is of course not historically factual. Setting aside such analysis, and speaking only as an untrained observer, I have long regarded this painting as presenting a symbolic image of the primordial state of scholarship.

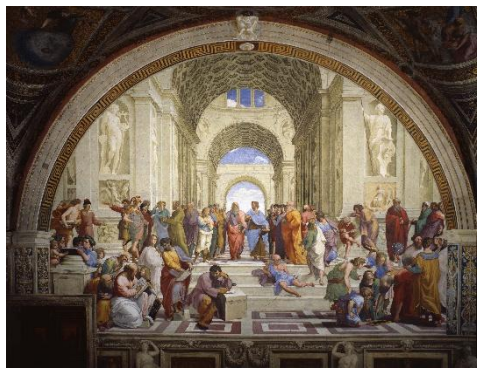

Figure 1. The School of Athens (Scuola di Atene)

In ancient times, the world was small. The learned minds of a city could gather in one place, and through direct dialogue and debate, the world of learning was largely complete. One who spoke on one day would listen on the next; listeners would become speakers. Ideas were exchanged freely, and science advanced through conversation.

In reality, civilization and culture did not belong to Greece alone, and as transportation and commerce expanded, so too did the world. Scholars from neighboring towns or regions could no longer meet daily, let alone those from other nations or continents. Instead of direct dialogue, thinkers wrote down their ideas and discoveries and sent them to distant colleagues. Some must have copied these texts by hand to spread knowledge more widely. The process was slow, but inevitable; as the world expanded, so too did the demands regarding the volume and speed of information dissemination.

Innovation arises in response to demand. In the mid-15th century, Johannes Gutenberg introduced movable-type printing in Europe and produced mass-printed Bibles. Employing this new technology, the world's first scholarly journals—*Journal des Sçavans* in France and *Philosophical Transactions* in the United Kingdom—were established in the mid-17th century, roughly two centuries later. Since then, scholarly journals have developed as a stable medium for distributing scientific knowledge globally.

Yet civilization did not stop evolving. The Industrial Revolution, followed in the 20th century by space exploration, energy research, and large-scale genomic science, brought forth fields of unprecedented scale. To support increasingly specialized academic domains, scholarly journals multiplied in number, frequency, and length. At the same time, the cost of information distribution became a critical issue.

This led to the next major innovation: electronic information dissemination.

By the late 20th century, the circulation of cutting-edge scientific information—research articles—had migrated to the internet in the form of electronic journals. To address cost issues from the author's side—that is, to deliver one's work to more readers by overcoming economic barriers—two additional methods emerged: self-archiving, and publishing in newly founded open access journals.

The open access movement quickly reached Japan. Over the ensuing two decades, nearly 800 domestic universities and research institutions have established institutional repositories (as of 2025). Publication in open access journals has expanded, and several universities have adopted so-called "read-and-publish" agreements that bundle subscription and open access publishing rights.

In retrospect, while the external form of scholarly communication has transformed dramatically, its essence has remained unchanged.

Electronic journals are but imitations of printed scholarly journals.

Disseminating one's work through institutional repositories is merely a reproduction of the traditional offprint-exchange culture.

This is unsurprising. It took two centuries for the innovation of printing technology to yield the killer application of scholarly journals. Even now, we have not yet produced a true killer application that fully exploits the potential of the internet. We are still in a process of trial and error—one that may again require two hundred years.

The League of European Research Universities (LERU) has stated that open science is a *culture change*. In Japan, however, open science is often explained narrowly as "open access to scholarly articles plus open access to research data." This definition is superficial and, in my view, inadequate. LERU describes open science as a mode of conducting research—a style of scholarly life—characterized by the following practices[i]:

1. *Make the resulting output, book or article, available as an Open Access output under an appropriate licence, ideally one of the Creative Commons licences.*
2. *Make the underlying research data, certainly the data used in the publication, available as an open dataset so that the conclusions reached in the publication can be checked and verified.*
3. *Make the research software, used for analysis, available so that the research is reproducible.*
4. *During the course of the research, consider making both the underlying research data and the publication available, the latter perhaps as a Green Open Access pre-print in a subject or institutional repository at each stage of the editing and review cycle prior to publication.*
5. *Of course, the activity in step 4 may not always be possible. For example, researchers may wish to retain primary use of their data until they have finished the round of publications which are to be based upon it. However, even in these cases, the actual processed data used in each publication could be made available as an open dataset. 6. In the publication and opening up of the supporting research data, it is highly desirable that a number of standard identifiers/processes be used to help discoverability and re-use of open outputs – ORCID to identify the authors; FundRef, a common taxonomy of research funder names; DOIs to identify and locate publications; DataCite to identify and locate datasets; Open Citations, a movement to promote the unrestricted availability of scholarly citation data, and to make these data available.*
6. *Any future killer application that fully harnesses digital information-distribution technologies must be capable of supporting this entire research lifecycle, including the process itself. To achieve this, such a system must be designed entirely independently from the inherent properties of print-based communication and the conventions, norms, and dogmas derived from it.*

A representative—and perhaps the most challenging—issue is breaking free from the *Version of Record* doctrine. A key difference between print-based publication (with typesetting, printing, binding, distribution) and online information dissemination is that the former cannot be corrected once produced. This characteristic led to the establishment of the final, fixed version of a paper—the Version of Record. Through peer review, a paper's value is established and crystallized into a single immutable version. Readers can then rely on this established content and share it confidently with the world. This stability has long been considered central to the scholarly journal's significance.

Yet was this truly a virtue?

Once printed, errors cannot be corrected—a clear flaw. In *The School of Athens*, would Plato never revise his words? Would Aristotle not change his thinking from one day to the next? We have perhaps glossed over the fundamental defect of print—the impossibility of revision—through positive reinterpretations such as those above, eventually convincing ourselves of our own rhetoric.

The concept of diamond open access—in which neither authors nor readers bear publication costs—is gaining global attention. Yet compared to subscription journals and author-pays open access journals, it represents no more than a change in business model. As noted earlier, the crucial issue is not how we open access, but how we reshape scholarly communication itself in a manner appropriate to the present and future. True evolution requires shedding the absolute authority of print-era conventions, embracing value transformation, and accepting the inherent modifiability of information. When a form of communication emerges that is both socially and technologically trustworthy, while grounded in electronic information technology, it will constitute a genuinely new killer application.

Research activity is moving toward global, real-time communication. Indeed, this is nothing new. The development of information-distribution technologies has always been driven by the aspiration to return to *The School of Athens*—an ideal realm in which science advances in real time.

As we observe the ongoing implementation of immediate open access policies[ii]—still rooted in self-archiving and open access journals, which are remnants of prior information-distribution modes—we must not neglect the parallel task of proactively developing diamond open access models suited to the internet era, and ultimately, of searching for entirely new forms of scholarly communication that lie beyond them.

---

[i] "Open Science and its role in universities: a roadmap for cultural change". League of European Research Universities. 2018. https://www.leru.org/publications/open-science-andits-role-in-universities-a-roadmap-for-cultural-change.

[ii] Basic Policy for Achieving Immediate Open Access to Academic Papers (Decision by Integrated Innovation Strategy Promotion Council, February 16) https://www8.cao.go.jp/cstp/oa_240216.pdf